\documentclass[preprint,12pt]{elsarticle}

\usepackage{epsfig}
\graphicspath{{figs/}}

\usepackage{amssymb}
\usepackage{amsmath}
\usepackage{bm}

\usepackage[ps2pdf,%
a4paper=true,%
breaklinks=true,%
colorlinks=true,%
pdfauthor={First Author et al.},%
pdftitle={Template for manuscripts in Advances in Space Research}%
]{hyperref}

\journal{Advances in Space Research}

\begin{document}

\begin{frontmatter}



\title{Suboptimal Formation Reconfiguration of Satellites Under Input Directional Constraints}


\author{Yasuhiro Yoshimura\corref{cor}}
\address{Tokyo Metropolitan University, 6-6 Asahigaoka, Hino, Tokyo}
\cortext[cor]{Corresponding author}
\ead{yyoshi@tmu.ac.jp}


\begin{abstract}
Proximity operations of satellites such as formation flying and on-orbit
servicing offer more advanced missions than missions achieved by
a single satellite. In a practical situation of formation flying,
thrust directions for keeping and controlling a relative orbit is
limited, e.g., for astronomical observation and plume impingement
avoidance. The aim of this paper is to provide an energy efficient
control method for a formation reconfiguration under input directional
constraints with respect to both an inertial and a leader-fixed frames.
The proposed controller is designed consisting of two parts: 1) guaranteeing
a formation reconfiguration to a desirable formation on the basis
of an energy optimal controller and 2) satisfying the input directional
constraints by superimposing additional inputs. The analytical form
of the control input shows that the input direction forms an ellipse
in the leader-fixed frame when a particular boundary condition is
satisfied, which is exploited as a nominal controller to take into
account the input directional constraints. Due to the singular avoidance
of the nominal controller, the additional inputs can be analytically
obtained. The effect on the follower trajectory due to the additional
inputs is compensated by setting a virtual target orbit, and thus
the successful formation reconfiguration is still guaranteed. Some
numerical simulation results verify the effectiveness of the proposed
method and compare the energy efficiency.\end{abstract}

\end{frontmatter}

\parindent=0.5 cm

\section{Introduction}

Relative motion control of satellites is a significant and promising
technology required for formation flying, on-orbit repairing and refuelling,
active debris removal, etc. In these missions, the precise and autonomous
control of satellite position and attitude with respect to target
objects is needed. For example, Project for On-Board Autonomy-3 (PROBA-3)
mission (\cite{Castellani:uk,SanchezMaestro:2013wh}), a formation flying
demonstration conducted by European Space Agency (ESA), requires a
relative positioning accuracy at the order of 1 mm for coronagraph
observation. Practical constraints on input directions also should
be considered in formation flying according to mission requirements.
A satellite formation for astronomical observation must orient its
telescope to a target object, which restricts large attitude change.
Thus, the thrust direction to keep or control a relative orbit is
limited with respect to an inertial frame. When a servicing satellite
maneuvers to a target satellite for inspection and repairing, the
thruster plume of the servicing satellite should not be oriented to
the target to avoid plume contamination, which indicates input directional
constraints with respect to the target satellite. 

Several studies have investigated formation control considering limited
input directions or input directional constraints. Most of studies
on formation control under limited input directions deal with along-track
thrusts because they can drive a satellite relative orbit to arbitrary
one, whereas cross-track thrusts cannot. \cite{Guelman:2011eb}
show a rendezvous method that uses along-track thrusts with bounded
magnitudes. \cite{Bando:2013kg} study an optimal
formation control by a single input in along-track direction and compare
the energy efficiency between continuous inputs and impulse ones.
On the other hand, \cite{Mitani:2011gc} investigate
a rendezvous method under input directional constraints with respect
to a target satellite. The controller in (\cite{Mitani:2011gc}) is
based on a satisficing method (\cite{Curtis:2004jj}) and can keep the
thrust direction not to be oriented to the target throughout the maneuver.
Furthermore an optimal controller in terms of $L_{1}$ and $L_{2}$
norms is shown in (\cite{Mitani:2013hi}) by using a smoothing method,
and an optimal formation control considering nonlinear relative equations
is proposed (\cite{Mitani:2014es}). In these subsequent analysis
and synthesis, the input directional constraint with respect to a
target satellite is studied, whereas \cite{Yoshimura:2013tu}
show an optimal reconfiguration method under an attitude constraint
with respect to an inertial frame. The control method in (\cite{Yoshimura:2013tu})
enables the simultaneous control of position and attitude of a satellite
that equips with only two thrusters. It is further shown that the
attitude constraint can be equivalently discussed as an input directional
constraint by tracking a reference input. Although the control method
shown in (\cite{Yoshimura:2013tu}) is applicable to formation control
under the input directional constraint with respect to an inertial
frame, it causes the oscillation of the input direction around a desired
direction. Thus, there still exists room for improvement of the control
method under input directional constraints with respect to the inertial
frame.

From a practical viewpoint, fuel or energy optimality is important
for keeping the relative motion of satellites in formation flying
because small satellite clusters are expected to achieve formation
flying missions (\cite{Castellani:uk,DelpechM:2013wh}). \cite{Carter:1996tf}
show an energy optimal formation control using upper and lower bounded
thrusts. \cite{Palmer:2006tt} analytically derives an energy
optimal reconfiguration method for formation flying in a circular
orbit by expressing control inputs with the Fourier series. This method
is further extended to a formation control in an elliptic orbit (\cite{Cho:2012ik})
and an optimal control problem under $J_{2}$ perturbation (\cite{Cho:2012if}).
Although these works have presented effective methods for designing
trajectories to desired relative orbits, much attention has not been
paid on input directional constraints.

In this context, the current study aims to present energy efficient
controller for formation flying under input directional constraints.
As discussed in (\cite{Yoshimura:2013tu}), formation control under
a constraint to an inertial frame is first derived by improving the
method in (\cite{Yoshimura:2013tu}). The proposed method in this paper
avoids the oscillation of thrusts around a target direction and is
further extended to incorporate with a constraint with respect to
a leader-fixed frame. These input directional constraints are satisfied
by superimposing additional inputs on a nominal input. Although the
control inputs are not strictly energy optimal due to the additional
inputs, the nominal controller is derived on the basis of an energy
optimal controller, which offers energy efficiency. 

This paper is organized as follows. Section 2 formulates the equations
of motion of satellites and the input directional constraints in formation
flying. In Section 3, an energy optimal controller that has no singularities
on a terminated time is shown, and then additional inputs for satisfying
the input directional constraints are obtained. Numerical simulation
results are shown in Section 4 to verify the effectiveness of the
proposed method, and Section 5 concludes this paper.

\section{Problem Formulation}

\subsection{Hill\textendash Clohessy\textendash Wiltshire equations}

\begin{figure}
\centering
\includegraphics[width=.6\textwidth]{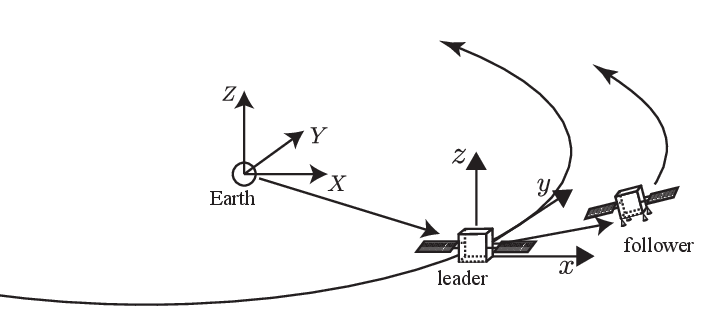}\protect\caption{Leader-fixed coordinate frame.}
\label{fig:leaderFixed}

\end{figure}

This study deals with the relative motion of two satellites in formation
flying. One satellite, called leader, is assumed to be orbiting in
a circular orbit, and the other satellite, called follower, is orbiting
in the proximity of the leader's orbit and is controlled to keep a
desired relative orbit with respect to the leader. The relative motion
of the follower satellite is described with respect to a leader-fixed
frame, in which three orthogonal axes are defined: $x$-axis directs
to a radial direction from the Earth, $z$-axis corresponds with the
direction of the orbital angular momentum vector of the leader, and
$y$-axis completes the right-handed coordinates (Fig. \ref{fig:leaderFixed}).
For simplicity, no disturbances on the satellites, such as the effect
of the Earth oblateness and solar radiation pressure, are assumed.
The equations of in-plane motion of the follower, called Hill\textendash Clohessy\textendash Wiltshire
(HCW) equations (\cite{Clohessy:1960kq}), are described as

\begin{align}
\frac{\mathrm{d}}{\mathrm{d}t}\left[\begin{array}{c}
x\\
\dot{x}\\
y\\
\dot{y}
\end{array}\right] & =\left[\begin{array}{cccc}
0 & 1 & 0 & 0\\
3\Omega & 0 & 0 & 2\Omega\\
0 & 0 & 0 & 1\\
0 & -2\Omega & 0 & 0
\end{array}\right]\left[\begin{array}{c}
x\\
\dot{x}\\
y\\
\dot{y}
\end{array}\right]\nonumber \\
 & +\left[\begin{array}{cc}
0 & 0\\
1 & 0\\
0 & 0\\
0 & 1
\end{array}\right]\left[\begin{array}{c}
u_{x}\\
u_{y}
\end{array}\right]\\
\Rightarrow & \dot{\bm{x}}=A_{{\rm HCW}}\bm{x}+B_{{\rm HCW}}\bm{u}
\end{align}
where the state variable vector $\bm{x}$ consists of the follower
position and velocity in the leader-fixed frame and $\Omega$ is the
orbital rate of the leader satellite. The control input $\bm{u}$
is assumed to be able to generate arbitrary accelerations. This paper
considers the relative in-plane motion alone because the cross-track
motion along the $z$-axis is a simple harmonic motion and is decoupled
from the in-plane motion. The analytical solution of Eq. (1) is described
as follows (\cite{Wie:1998tga}). 
\begin{align}
&\bm{x}\left(t\right) =\left[\begin{array}{cccc}
4-3c_{t} & s_{t}/\Omega & 0 & 2\left( 1-c_{t}\right) /\Omega\\
3\Omega s_{t} & c_{t} & 0 & 2s_{t}\\
6\left[ s_{t}-\Omega\left(t-t_{0}\right)\right]  & -2\left( 1-c_{t}\right) /\Omega & 1 & -3\left(t-t_{0}\right)+4s_{t}/\Omega\\
-6\Omega\left( 1-c_{t}\right)  & -2s_{t} & 0 & 4c_{t}-3
\end{array}\right]\bm{x}\left(t_{0}\right) \nonumber\\
&+\int_{t_{0}}^{t}B\bm{u}\mathrm{d}t\\
&= \Phi\left(t,t_{0}\right)\bm{x}\left(t_{0}\right)+\int_{t_{0}}^{t}\Phi\left(t, \tau\right)B\bm{u}\mathrm{d}\tau\label{eq:HCWsolution}
\end{align}
where $c_{t}:=\cos\left[\Omega\left(t-t_{0}\right)\right]$, $s_{t}:=\sin\left[\Omega\left(t-t_{0}\right)\right]$, and $t_{f}$ and $t_{0}$ are a terminated and initial time of a
reconfiguration maneuver, respectively. The matrix $\Phi\left(t,t_{0}\right)$
is the state transition matrix connecting the states $\bm{x}\left(t_{0}\right)$
and $\bm{x}\left(t\right)$ with no inputs, i.e., $\bm{x}\left(t_{f}\right)=\Phi\left(t,\, t_{0}\right)\bm{x}\left(t_{0}\right)$.
The state transition matrix consists of a fundamental matrix $\Phi_{xy}$
as
\begin{equation}
\Phi\left(t,t_{0}\right)=\Phi_{xy}\left(t\right)\Phi_{xy}^{-1}\left(t_{0}\right)
\end{equation}
where
\begin{equation}
\Phi_{xy}\left(t\right)=\left[\begin{array}{cccc}
0 & -\sin\left(\Omega t\right)/\Omega & -\cos\left(\Omega t\right) & -2/\Omega\\
0 & -\cos\left(\Omega t\right) & \Omega\sin\left(\Omega t\right) & 0\\
1 & -2\cos\left(\Omega t\right)/\Omega & 2\sin\left(\Omega t\right) & 3t\\
0 & 2\sin\left(\Omega t\right) & 2\Omega\cos\left(\Omega t\right) & 3
\end{array}\right]
\end{equation}

The analytical solution of HCW equations without external forces is rewritten as
\begin{align}
x\left(t\right) & =-a\cos\left(\Omega t+\alpha\right)+\frac{2b}{\Omega}\label{eq:solutionX}\\
y\left(t\right) & =2a\sin\left(\Omega t+\alpha\right)-3bt+d\label{eq:solutionY}
\end{align}
where
\begin{align}
a & :=\sqrt{\left[3x\left(t_{0}\right)+2\frac{\dot{y}\left(t_{0}\right)}{\Omega}\right]^{2}+\left[\frac{\dot{x}\left(t_{0}\right)}{\Omega}\right]^{2}}\\
b & :=2\Omega x\left(t_{0}\right)+\dot{y}\left(t_{0}\right)\\
d & :=y\left(t_{0}\right)-\frac{2\dot{x}\left(t_{0}\right)}{\Omega}\\
\cos\alpha & =\frac{\left[3x\left(t_{0}\right)+2\dot{y}\left(t_{0}\right)/\Omega\right]}{a}\\
\sin\alpha & =\frac{\dot{x}\left(t_{0}\right)}{a\Omega}
\end{align}
Equations~\eqref{eq:solutionX} and~\eqref{eq:solutionY} show that
the follower satellite relatively flies around the leader when $b=0$.
The relative orbit is ellipse with the ratio $\left|x\left(t\right)\right|/\left|y\left(t\right)\right|=0.5$
and its center position is written as $\left(x,\, y\right)=\left(d,\,0\right)$.
This paper considers a formation reconfiguration and assumes that
the follower initially fly around the leader satellite, i.e., $d=b=0$.

\subsection{Input directional constraints}

In formation flying, input directional constraints should be considered
according to mission requirements. For instance, an input directional
constraint with respect to an inertial frame stems from mission requirements
for astronomical observation or space solar power system.
The limitation of the attitude change due to the power generation with fixed solar arrays restricts
the thrust direction with respect to the inertial frame. The follower
satellite thus needs to control a relative orbit while keeping the
thrust direction along a specified direction in the inertial frame. 

For redundancy, many thrusters are equipped on a satellite so that they can generate translational forces and rotational torques in arbitrary directions.
This indicates that translational forces in the $x$-$y$ plane can be generated by properly distributing the magnitude of each thrust.
Thus this study considers thruster number and  relative motion of satellites and thruster forces in the $x$-$y$ plane without loss of generality.

\begin{figure}
\centering
\includegraphics[width=.9\textwidth]{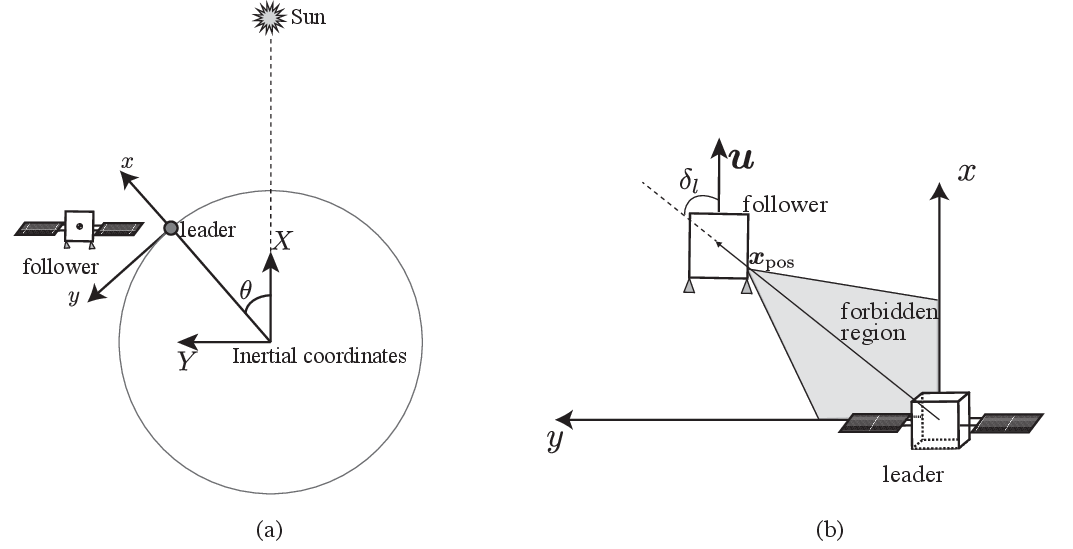}\protect\caption{Input directional constraints with respect to (a) an inertial frame,
and to (b) a leader-fixed frame.}
\label{fig:constraints}

\end{figure}

Since the inertial frame can be set arbitrarily, without loss of generality,
this paper assumes that the desired direction is the positive direction
of the $X$-axis in the inertial frame as described in Fig. \ref{fig:constraints}(a).
In other words, by defining the thrust directional angle with respect
to the inertial frame as
\begin{equation}
\delta_{i}:=\arctan\left(\frac{u_{y}}{u_{x}}\right)+\theta\label{eq:deltaI}
\end{equation}
the input directional constraint denotes that the angle $\delta_{i}$
should be kept zero throughout the maneuver. In Eq.~\eqref{eq:deltaI},
$\theta$ is the true anomaly of the leader satellite.

In addition, this study investigates a formation control under the
input directional constraint with respect to the leader satellite.
Such constraint arises in an autonomous rendezvous and on-orbit servicing
mission for avoiding plume impingement and/or contamination from the
follower\textquoteright s thrusts to the leader. The thrust direction
of the follower with respect to the leader is limited, and the follower
must control the relative orbit to a desired orbit while keeping the
thrust vector not to orient into the leader as shown in Fig. \ref{fig:constraints}(b). 

The angle between the position vector of the follower $\bm{x}_{{\rm pos}}=\left[x,y\right]^{T}$
and the acceleration vector $\bm{u}$ is defined as 
\begin{equation}
\delta_{l}:=\arccos\left(\frac{\bm{u}\cdot\bm{x}_{{\rm pos}}}{\left|\bm{u}\right|\left|\bm{x}_{{\rm pos}}\right|}\right)
\end{equation}
Since the thruster plume is oriented into $-\bm{u}$, the forbidden
region of the thruster plume is described as $\left|\delta_{l}\right|\leq\bar{\delta}$,
where $\bar{\delta}$ is an upper bound angle. The input directional
constraint with respect to the leader satellite requires that the
angle $\delta_{l}$ should avoid being within the angle $\bar{\delta}$
throughout the maneuver. 

\section{Control Method}

This section derives a suboptimal reconfiguration method consisting
of two parts: the first part is an energy optimal controller for guaranteeing
the formation reconfiguration to a desired relative orbit and the
second part is an additional controller that is superimposed onto
the optimal controller for satisfying the input directional constraints.
Though the additional controller makes the follower position different
from the optimal one, such undesirable displacements are compensated
by setting a virtual target orbit.

\subsection{Energy optimal controller}

The energy optimal controller in this paper is based on the method
proposed by \cite{Palmer:2006tt}. The controller in (\cite{Palmer:2006tt})
uses the transformation of boundary conditions to facilitate the derivation.
Such transformation causes a singularity that disables setting a maneuver
terminated time $t_{f}$ as the multiple of the orbital period, i.e.,
setting $t_{f}=2\pi N/\Omega$ $(N=1,2,\dots)$ is impossible. This
singularity can be avoided by formulating the optimal controller without the transformation (\cite{Cho:2012ik}). The advantage of the singularity
avoidance is that setting $t_{f}=2\pi N/\Omega$ simplifies the calculations
for satisfying the input directional constraints, as shown in later.

An energy optimal controller is designed for minimizing the following
cost function: 
\begin{equation}
J=\int_{0}^{t_{f}}\left(u_{x}^{2}+u_{y}^{2}\right)\mathrm{d}t\label{eq:costFunction}
\end{equation}
In Eq.~\eqref{eq:costFunction} and the following discussions, the
initial time is set to $t_{0}=0$ for simplicity. The control input
is described with the Fourier series for expressing arbitrary inputs
as
\begin{align}
u_{x} & =\frac{a_{x0}}{2}+\sum_{n=1}^{\infty}\left[a_{xn}\cos\left(\frac{2n\pi}{t_{f}}t\right)+b_{xn}\sin\left(\frac{2n\pi}{t_{f}}t\right)\right]\label{eq:uxFourier}\\
u_{y} & =\frac{a_{y0}}{2}+\sum_{n=1}^{\infty}\left[a_{yn}\cos\left(\frac{2n\pi}{t_{f}}t\right)+b_{yn}\sin\left(\frac{2n\pi}{t_{f}}t\right)\right]\label{eq:uyFourier}
\end{align}
where $a_{j0,}\, a_{jn},\, b_{jn}\,\left(j=x,y\right)$ are the Fourier
coefficients. By substituting Eqs.~\eqref{eq:uxFourier} and~\eqref{eq:uyFourier}
into Eq.~\eqref{eq:costFunction} and using the Parseval's theorem (\cite{Boas:2006ty}),
the cost function in Eq.~\eqref{eq:costFunction} is rewritten as
follows.
\begin{align}
J & =\frac{t_{f}}{2}\left[\frac{a_{x0}^{2}}{2}+\sum_{n=1}^{\infty}\left(a_{xn}^{2}+b_{xn}^{2}\right)\right]\nonumber \\
 & +\frac{t_{f}}{2}\left[\frac{a_{y0}^{2}}{2}+\sum_{n=1}^{\infty}\left(a_{yn}^{2}+b_{yn}^{2}\right)\right]
\end{align}

The relative orbit control must satisfy a desirable state $\bm{x}_{d}$
at the terminated time $t_{f}$. From the analytical solution of the
HCW equations in Eq.~\eqref{eq:HCWsolution}, the follower state at
the terminated time is written as
\begin{equation}
\bm{x}\left(t_{f}\right)=\bm{x}_{h}+\Phi_{xy}\left(t_{f}\right)\int_{0}^{t_{f}}\Phi_{xy}^{-1}\left(\tau\right)B\bm{u}\mathrm{d}\tau
\end{equation}
where $\bm{x}_{h}$ is the homogeneous solution of the HCW equations.
The boundary condition vector $\bm{Q}$ is defined as follows.
\begin{align}
\bm{Q}: & =\Phi_{xy}^{-1}\left(t_{f}\right)\left[\bm{x}\left(t_{f}\right)-\bm{x}_{h}\right]\label{eq:givenQ}\\
= & \int_{0}^{t_{f}}\Phi_{xy}^{-1}\left(\tau\right)B\bm{u}\mathrm{d}\tau\label{eq:givenQ2}
\end{align}
The integrand in Eq.~\eqref{eq:givenQ2} is calculated as
\begin{equation}
\Phi_{xy}^{-1}\left(\tau\right)B\bm{u}=\left[\begin{array}{c}
-\frac{2u_{x}}{\Omega}+3\tau u_{y}\\
-u_{x}\cos\left(\Omega\tau\right)+2u_{y}\sin\left(\Omega\tau\right)\\
\frac{u_{x}}{\Omega}\sin\left(\Omega\tau\right)+2u_{y}\Omega\cos\left(\Omega\tau\right)\\
-u_{y}
\end{array}\right]\label{eq:integrand}
\end{equation}
The substitution of Eqs.~\eqref{eq:uxFourier},~\eqref{eq:uyFourier},
and~\eqref{eq:integrand} into Eq.~\eqref{eq:givenQ2} yields
\begin{align}
\int_{0}^{t_{f}}\Phi_{xy}^{-1}\left(\tau\right)B\bm{u}\mathrm{d}\tau & =\left[\begin{array}{c}
\frac{j_{0}}{2}a_{x0}+\sum_{n=1}^{\infty}\left(j_{a}a_{xn}+j_{b}b_{xn}\right)+\frac{l_{0}}{2}a_{y0}+\sum_{n=1}^{\infty}\left(l_{a}a_{yn}+l_{b}b_{yn}\right)\\
\frac{h_{0}}{2}a_{x0}+\sum_{n=1}^{\infty}\left(h_{a}a_{xn}+h_{b}b_{xn}\right)+\frac{k_{0}}{2}a_{y0}+\sum_{n=1}^{\infty}\left(k_{a}a_{yn}+k_{b}b_{yn}\right)\\
\frac{p_{0}}{2}a_{x0}+\sum_{n=1}^{\infty}\left(p_{a}a_{xn}+p_{b}b_{xn}\right)+\frac{m_{0}}{2}a_{y0}+\sum_{n=1}^{\infty}\left(m_{a}a_{yn}+m_{b}b_{yn}\right)\\
\frac{q_{0}}{2}a_{y0}+\sum_{n=1}^{\infty}\left(q_{a}a_{yn}+q_{b}b_{yn}\right)
\end{array}\right]\label{eq:integral}
\end{align}
The definitions of the variables in Eq.~\eqref{eq:integral} are summarized
in Appendix.

The optimal control problem is finding the control inputs $\bm{u}$
that minimizes the cost function in Eq.~\eqref{eq:costFunction} and
satisfies Eq.~\eqref{eq:givenQ} that is calculated with given boundary
states and terminated time. Furthermore, since the control inputs
are expressed with the Fourier series as shown in Eqs.~\eqref{eq:uxFourier}
and~\eqref{eq:uyFourier}, the optimal control problem is reduced
to finding the Fourier coefficients. 

The Hamiltonian for minimizing the cost function~\eqref{eq:costFunction}
is described as 
\begin{align}
H= & \ensuremath{\frac{{t_{f}}}{2}\left[{\frac{{a_{x0}^{2}}}{2}+\sum\limits _{n=1}^{\infty}{\left({a_{xn}^{2}+b_{xn}^{2}}\right)}}\right]+\frac{{t_{f}}}{2}\left[{\frac{{a_{y0}^{2}}}{2}+\sum\limits _{n=1}^{\infty}{\left({a_{yn}^{2}+b_{yn}^{2}}\right)}}\right]}\nonumber \\
 & +{{\bm{\lambda}}^{T}}\left[\bm{Q}-\int_{t_{0}}^{t_{f}}\Phi_{xy}^{-1}(\tau)B{\bm{u}}{\text{d}}\tau\right]\label{eq:Hamiltonian}
\end{align}
where $\bm{\lambda}=\left[\lambda_{1},\lambda_{2},\lambda_{3},\lambda_{4}\right]^{T}$
denotes the Lagrange multipliers. Substituting Eqs.~\eqref{eq:uxFourier}
and~\eqref{eq:uyFourier} into Eq.~\eqref{eq:Hamiltonian} and the
differentiating the Hamiltonian with respect to each Fourier coefficient
yield the following equations.
\begin{align}
\frac{t_{f}}{2}a_{x0}-\left(\frac{j_{0}}{2}\lambda_{1}+\frac{h_{0}}{2}\lambda_{2}+\frac{p_{0}}{2}\lambda_{3}\right) & =0\label{eq:dax0}\\
\sum_{n=1}^{\infty}\left(t_{f}a_{xn}\right)-\sum_{n=1}^{\infty}\left(j_{a}\lambda_{1}+h_{a}\lambda_{2}+p_{a}\lambda_{3}\right) & =0\label{eq:daxn}\\
\sum_{n=1}^{\infty}\left(t_{f}b_{xn}\right)-\sum_{n=1}^{\infty}\left(j_{b}\lambda_{1}+h_{b}\lambda_{2}+p_{b}\lambda_{3}\right) & =0\label{eq:dbxn}\\
\frac{t_{f}}{2}a_{y0}-\left(\frac{l_{0}}{2}\lambda_{1}+\frac{k_{0}}{2}\lambda_{2}+\frac{m_{0}}{2}\lambda_{3}+\frac{q_{0}}{2}\lambda_{4}\right) & =0\label{eq:day0}\\
\sum_{n=1}^{\infty}\left(t_{f}a_{yn}\right)-\sum_{n=1}^{\infty}\left(l_{a}\lambda_{1}+k_{a}\lambda_{2}+m_{a}\lambda_{3}+q_{a}\lambda_{4}\right) & =0\label{eq:dayn}\\
\sum_{n=1}^{\infty}\left(t_{f}b_{yn}\right)-\sum_{n=1}^{\infty}\left(l_{b}\lambda_{1}+k_{b}\lambda_{2}+m_{b}\lambda_{3}+q_{b}\lambda_{4}\right) & =0\label{eq:dbyn}
\end{align}
 From Eqs.~\eqref{eq:dax0}\textendash~\eqref{eq:dbyn}, the Fourier
coefficients are expressed with the Lagrange multipliers as follows.

\begin{align}
a_{x0} & =\frac{2}{t_{f}}\left(\frac{j_{0}}{2}\lambda_{1}+\frac{h_{0}}{2}\lambda_{2}+\frac{p_{0}}{2}\lambda_{3}\right)\label{eq:ax0}\\
\sum_{n=1}^{\infty}a_{xn} & =\frac{1}{t_{f}}\sum_{n=1}^{\infty}\left(j_{a}\lambda_{1}+h_{a}\lambda_{2}+p_{a}\lambda_{3}\right)\label{eq:axn}\\
\sum_{n=1}^{\infty}b_{xn} & =\frac{1}{t_{f}}\sum_{n=1}^{\infty}\left(j_{b}\lambda_{1}+h_{b}\lambda_{2}+p_{b}\lambda_{3}\right)\label{eq:bxn}\\
a_{y0} & =\frac{2}{t_{f}}\left(l_{0}\lambda_{1}+k_{0}\lambda_{2}+m_{0}\lambda_{3}+q_{0}\lambda_{4}\right)\label{eq:ay0}\\
\sum_{n=1}^{\infty}a_{yn} & =\frac{1}{t_{f}}\sum_{n=1}^{\infty}\left(l_{a}\lambda_{1}+k_{a}\lambda_{2}+m_{a}\lambda_{3}+q_{a}\lambda_{4}\right)\label{eq:ayn}\\
\sum_{n=1}^{\infty}b_{yn} & =\frac{1}{t_{f}}\sum_{n=1}^{\infty}\left(l_{b}\lambda_{1}+k_{b}\lambda_{2}+m_{b}\lambda_{3}+q_{b}\lambda_{4}\right)\label{eq:byn}
\end{align}
Substituting these Fourier coefficients into Eq.~\eqref{eq:givenQ2}
and using the Parseval's theorem, we obtain the following equation.
\begin{align}
\bm{Q} & =A\bm{\lambda}\nonumber \\
= & \left[\begin{array}{cccc}
A_{11} & A_{12} & A_{13} & A_{14}\\
A_{12} & A_{22} & A_{23} & A_{24}\\
A_{13} & A_{23} & A_{33} & A_{34}\\
A_{14} & A_{24} & A_{34} & A_{44}
\end{array}\right]\left[\begin{array}{c}
\lambda_{1}\\
\lambda_{2}\\
\lambda_{3}\\
\lambda_{4}
\end{array}\right]\label{eq:Q}
\end{align}
This algebraic equation can be analytically solved for the Lagrange
multipliers $\bm{\lambda}$ with arbitrary initial and desired states
and terminated time (see Appendix for the components of the matrix
$A$). 

By substituting the multipliers back into Eqs.~\eqref{eq:ax0}\textendash~\eqref{eq:byn},
we obtain the analytical forms of the Fourier coefficients, which
are equivalent to the optimal control inputs:
\begin{align}
u_{x} & =-\frac{2}{\Omega}\lambda_{1}-\lambda_{2}\cos\left(\Omega t\right)+\frac{1}{\Omega}\lambda_{3}\sin\left(\Omega t\right)\\
u_{y} & =3\Omega\lambda_{1}t+2\lambda_{2}\sin\left(\Omega t\right)+\frac{2}{\Omega}\lambda_{3}\cos\left(\Omega t\right)-\lambda_{4}
\end{align}
 or alternatively,
\begin{align}
u_{x} & =-\frac{2}{\Omega}\lambda_{1}+\Lambda\sin\left(\Omega t-\Psi\right)\label{eq:ux}\\
u_{y} & =3\Omega\lambda_{1}t+2\Lambda\cos\left(\Omega t-\Psi\right)-\lambda_{4}\label{eq:uy}
\end{align}
where
\begin{align}
\Lambda & =\sqrt{\lambda_{2}^{2}+\left(\lambda_{3}/\Omega\right)^{2}}\label{eq:Lambda}\\
\tan\Psi & =\Omega\lambda_{2}/\lambda_{3}
\end{align}

Note that, although the optimal controller in Eqs.~\eqref{eq:ux}
and~\eqref{eq:uy} has the same form shown in (\cite{Palmer:2006tt}),
the derived controller in the current paper has no singularity, that
is, the terminated time can be set to $t_{f}=2\pi N/\Omega$. Such
terminated time can simplify further calculations for satisfying the
input directional constraints as shown in the following subsections.

\subsection{Additional input for the constraint with respect to the inertial
frame}

Since the desired direction, the $X$-axis, is the axis in the inertial frame, the $X$-axis expressed in the leader-fixed
frame is not constant but monotonically varying due to the orbital
motion of the leader satellite as $\dot{\psi}_{d}=-\Omega$. If the
input direction with respect to the leader-fixed frame also varies
monotonically, the constraint on the input direction is expected to
be satisfied, which is the control technique used in (\cite{Yoshimura:2013tu}).
In fact, the input direction can be designed for changing in one
direction, because the derived controller in Eqs.~\eqref{eq:ux} and~\eqref{eq:uy} forms an ellipse when $\lambda_{1}=0$ and furthermore
an origin-centered ellipse when $\lambda_{1}=\lambda_{4}=0$. The
drawback is that, the time derivative of the input angle is not constant, whereas that of the desired direction
is constant. Such difference causes the oscillation of the input direction
around the desired one (\cite{Yoshimura:2013tu}). To solve this problem,
an additional input that orients the input direction at the same rate
as the orbital rate is used in this paper. 

The boundary condition for $\lambda_{1}=0$ has been previously derived
by \cite{Yoshimura:2013tu} as follows.
\begin{align}
\frac{y\left(0\right)}{2x\left(0\right)} & =\tan\left(\frac{\Omega t_{f}}{2}\right)\label{eq:initialCondition}\\
\frac{y\left(t_{f}\right)}{2x\left(t_{f}\right)} & =\tan\left(-\frac{\Omega t_{f}}{2}\right)\label{eq:terminatedCondition}
\end{align}
Starting and terminating the reconfiguration maneuver at these boundary
states realize $\lambda_{1}=0$, resulting in an elliptic input direction
as:
\begin{align}
u_{x} & =\Lambda\sin\left(\Omega t-\Psi\right)\\
u_{y} & =2\Lambda\cos\left(\Omega t-\Psi\right)-\lambda_{4}
\end{align}
Under the assumption that $\lambda_{4}\approx0$, the input angle
in the leader-fixed frame and its derivative are described as 
\begin{align}
\psi: & =\arctan\left(\frac{u_{y}}{u_{x}}\right)\label{eq:psi111}\\
\dot{\psi} & =-\frac{2\Omega}{4-3\sin\left(\Omega t-\Psi\right)}
\end{align}
Since $\dot{\psi}_{d}=-\Omega$, the input angle does not correspond
with the desired direction throughout the maneuver even if the initial
directions coincide. It is noted that, although the multiplier $\lambda_{4}$
is not exactly zero, the larger $N$ is, the smaller the multiplier
$\lambda_{4}$ becomes as seen in the 4th row of Eq.~\eqref{eq:LagrangeMultipliers}.
That is, setting long terminated time $t_{f}$$\left(=2N\pi/\Omega\right)$
makes the assumption $\lambda_{4}\approx0$ reasonable.

To orient the input direction at the same rate as the desired direction,
the following additional input $\bm{u}_{{\rm add},i}$ is superimposed
onto the nominal optimal input $\bm{u}$.
\begin{equation}
\bm{u}_{{\rm add},i}=\left[\begin{array}{c}
\Lambda\sin\left(\Omega t-\Psi\right)\\
0
\end{array}\right]\label{eq:uAddI}
\end{equation}
The resulting input is described as
\begin{align}
u_{x} & =2\Lambda\sin\left(\Omega t-\Psi\right)\label{eq:uxModified}\\
u_{y} & =2\Lambda\cos\left(\Omega t-\Psi\right)-\lambda_{4}\label{eq:uyModified}
\end{align}
This additional input can make the input direction circular, which
has the same rate as the orbital rate of the leader. In fact, the
time derivative of the input angle is calculated as $\dot{\psi}=-\Omega$
using Eqs.~\eqref{eq:uxModified} and~\eqref{eq:uyModified}.

The additional input affects the transition of the follower to the
target orbit and causes an error. Due to this error, the relative orbit of the follower does not converge to a relative target orbit.
Thus this undesirable error is compensated
by setting a virtual target orbit. The effect of the additional input
to the follower position can be calculated as follows.
\begin{align}
\Delta\bm{x} & =\int_{0}^{t_{f}}\Phi\left(t_{f}, \tau\right)B\bm{u}_{{\rm add},i}\mathrm{d}\tau\nonumber \\
= & \left[\begin{array}{c}
-\frac{\Lambda\left[t_{f}\Omega\cos\left(\Omega t_{f}-\Psi\right)-\cos\Psi\sin\left(\Omega t_{f}\right)\right]}{2\Omega^{2}}\\
-\frac{\Lambda\left(t_{f}\Omega\cos\left(\Omega t_{f}\right)\sin\Psi+\left(-t_{f}\Omega\cos\Psi+\sin\Psi\right)\sin\left(\Omega t_{f}\right)\right)}{2\Omega}\\
\frac{\Lambda\left[-4\cos\Psi+3\cos\left(\Omega t_{f}-\Psi\right)+\cos\left(\Omega t_{f}+\Psi\right)+2\Omega t_{f}\sin\left(\Omega t_{f}-\Psi\right)\right]}{2\Omega^{2}}\\
\Lambda t_{f}\cos\left(\Omega t_{f}-\Psi\right)-\frac{\Lambda\cos\Psi\sin\left(\Omega t_{f}\right)}{\Omega}
\end{array}\right]\label{eq:deltaXi}
\end{align}
Although the effect of the additional input is analytically described in Eq.~\eqref{eq:deltaXi}, the convergence of the follower to the target relative orbit is not guaranteed due to the additional input.
Thus the state variation due to the additional inputs should be reduced.

Setting the terminated time $t_{f}=2\pi N/\Omega$ can  simplify
Eq.~\eqref{eq:deltaXi} as 
\begin{equation}
\Delta\bm{x}=\left[\begin{array}{c}
-\frac{\pi\Lambda}{\Omega^{2}}\\
0\\
0\\
\frac{2\pi\Lambda}{\Omega}
\end{array}\right]\label{eq:Deltax}
\end{equation}
From Eqs.~\eqref{eq:givenQ} and~\eqref{eq:Q}, the Lagrange multipliers can be easily calculated using MAPLE or Mathematica as follows:
\begin{equation}
\bm{\lambda} = \left[ {\begin{array}{*{20}{c}}
  0 \\ 
  0 \\ 
  { - \frac{{{\Omega ^3}\left[ {x({t_f}) - {x_h}} \right]}}{{5N\pi }}} \\ 
  0 
\end{array}} \right]
\end{equation}
Then, the variable $\Lambda$ is also obtained from Eq.~\eqref{eq:Lambda} as 
\begin{equation}
\Lambda=\frac{\Omega^{2}\left[x\left(t_{f}\right)-x_{h}\right]}{{5N\pi}}
\end{equation}
Substituting $\Lambda$ into the first component of Eq.~\eqref{eq:Deltax}
yields the virtual target position along the $x$-axis as
\begin{equation}
x_{vd}=\frac{5}{6}x_{d}-\frac{1}{6}x\left(0\right) \label{eq:xvd}
\end{equation}
where $x_{d}$ is the target position along the $x$-axis. The formation
reconfiguration from $\bm{x}\left(0\right)$ to $\bm{x}_{d}$ is achieved
using the suboptimal input in Eqs.~\eqref{eq:uxModified} and~\eqref{eq:uyModified},
in which the virtual target orbit is used.

The proposed control method under the constraint with respect to the inertial frame is summarized as follows.
The terminated time of the maneuver $t_{f}$ is determined as the multiple of the orbital period.
Then, substituting $t_{f}$ into Eqs.~\eqref{eq:initialCondition} and~\eqref{eq:terminatedCondition} provides the maneuver starting position and terminated one of the follower.
The control input shown in Eqs.~\eqref{eq:uxModified} and~\eqref{eq:uyModified} is implemented, in which the Lagrange multipliers are calculated using the
virtual target position along the $x$-axis in Eq.~\eqref{eq:xvd}

\subsection{Additional inputs for the constraint with respect to the leader-fixed
frame}

This subsection extends the control method to accommodate the input
directional constraint with respect to the leader-fixed frame. Since
the follower satellite is relatively orbiting around the leader, the
forbidden region of the thrust direction monotonically varies. Thus
the input direction that changes elliptically is exploited also to keep the input
direction within the desired one in the leader-fixed frame.

The initial input direction must be in the admissible region to satisfy
the input directional constraint throughout the maneuver. From Eqs.~\eqref{eq:ux} and~\eqref{eq:uy} and $\lambda_{1}=0$, the initial
input angle is expressed with
\begin{align}
\psi\left(0\right) & =\arctan\left[\frac{u_{y}\left(0\right)}{u_{x}\left(0\right)}\right]\\
= & \arctan\left[\frac{2\Lambda\cos\left(-\Psi\right)-\lambda_{4}}{\Lambda\sin\left(-\Psi\right)}\right]\label{eq:psi0}
\end{align}
The input angle $\psi\left(0\right)$ is a function of the Lagrange
multipliers, and furthermore the multipliers are the function of the
terminated time $t_{f}$, making $\psi\left(0\right)$ highly nonlinear.
Nevertheless the initial input direction can be simplified by setting
the terminated time as $t_{f}=2\pi N/\Omega$, which is the reason
that the optimal controller without singularities is derived. Substituting
$t_{f}=2N\pi/\Omega$ into Eq.~\eqref{eq:initialCondition} yields
\begin{align}
\frac{y\left(0\right)}{2x\left(0\right)} & =0\\
\Rightarrow & y\left(0\right)=0\label{eq:initialY}
\end{align}
The initial input angle $\psi\left(0\right)$ is then determined as
\begin{equation}
\psi\left(0\right)=\pi/2\label{eq:initialPsi}
\end{equation}
Therefore the initial thrust direction $-\bm{u}$ is analytically
obtained and is orthogonal to the follower position vector regardless
of the initial and target orbit.

Both the follower trajectory and the input direction vary elliptically.
However, the maneuver starts when the follower is on the $x$-axis as derived
in Eq.~\eqref{eq:initialY}, whereas the initial input angle is $\pi/2$
as shown in Eq.~\eqref{eq:initialPsi}. Thus the initial phase angle differs
by 90 deg and this different phase causes an error. 
To eliminate this phase difference, an additional
input is superimposed onto the nominal inputs described in Eqs.~\eqref{eq:ux}
and~\eqref{eq:uy}. 

Since the nominal input direction becomes ellipse with the ratio
$\left|u_{x}\right|/\left|u_{y}\right|=1/2$, the additional input
making the ratio $\left|u_{x}\right|/\left|u_{y}\right|=2$ is required
for compensating the phase difference. The following additional input
for the acceleration along the $y$-axis is added to the nominal input
$\bm{u}$.
\begin{equation}
\bm{u}_{{\rm add},l}=\left[\begin{array}{c}
0\\
-\frac{3}{2}\Lambda\cos\left(\Omega t-\Psi\right)
\end{array}\right]
\end{equation}
This additional input can make the input direction ellipse at the
ratio $\left|u_{x}\right|/\left|u_{y}\right|=2$, resulting in the
same phase as the formation reconfiguration of the follower. 

In the similar manner to the previous subsection, the effect of the
additional input to the follower position can be calculated as follows.
\begin{align}
\Delta\bm{x} & =\int_{t_{0}}^{t_{f}}\Phi\left(t_{f}, \tau\right)B\bm{u}_{{\rm add},l}\mathrm{d}\tau\\
= & \left[\begin{array}{c}
\frac{3\pi\Lambda}{\Omega^{2}}\\
0\\
0\\
-\frac{6\pi\Lambda}{\Omega}
\end{array}\right]\label{eq:deltaX2}
\end{align}
It is noted that the position variation $\Delta\bm{x}$ in Eq.~\eqref{eq:deltaX2}
is also simplified by setting the terminated time $t_{f}=2\pi N/\Omega$
, whereas using the other terminated time results in a more complicated
form. The virtual target orbit to compensate the variation in Eq.~\eqref{eq:deltaX2} can be analytically obtained as
\begin{equation}
x_{vd}=\frac{5}{2}x_{d}-\frac{3}{2}x\left(0\right)\label{eq:xvdLeader}
\end{equation}

The proposed control method under the constraint with respect to the leader-fixed frame is described as follows.
The terminated time of the maneuver $t_{f}$ is determined as the multiple of the orbital period.
The maneuver starting and terminated positions are obtained by substituting $t_{f}$ into Eqs.~\eqref{eq:initialCondition} and~\eqref{eq:terminatedCondition}.
In the maneuver, the control input $\bm{u}+\bm{u}_{{\rm add,}l}$ is implemented, in which the Lagrange multipliers are calculated using the
virtual target position along the $x$-axis in Eq.~\eqref{eq:xvdLeader}.

\section{Numerical simulation results}

This section demonstrates numerical simulation results to verify the
effectiveness of the proposed formation controller. The simulations
are performed setting the initial relative semi-major axis to 3000
m and the target one to 1000 m. The leader satellite is assumed to
be orbiting at the rate of $\Omega=6.31\times10^{-4}\,{\rm rad/s}$
in a circular orbit, which corresponds to 10000 km semi-major axis. 

\begin{figure}
\centering
\includegraphics[width=.5\textwidth]{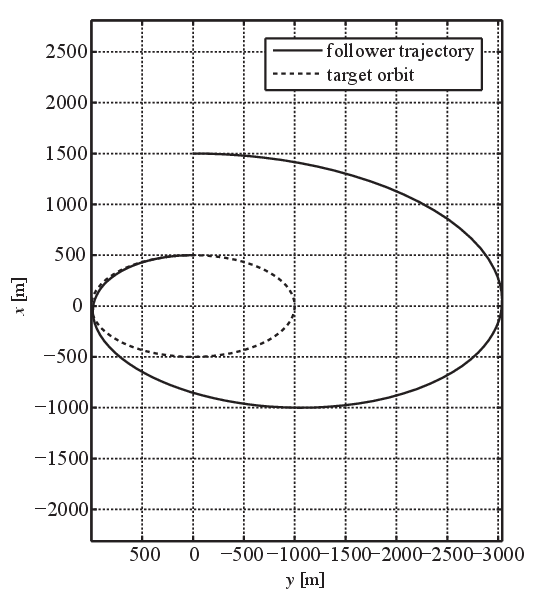}

\protect\caption{Follower trajectory under the constraint on the input direction with
respect to the inertial frame.}
\label{fig:01trajectory}
\end{figure}

\begin{figure}
\centering
\includegraphics[width=.5\textwidth]{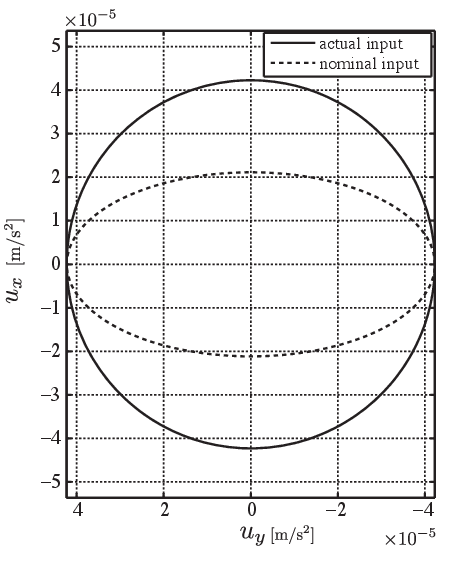}\protect\caption{Input directions of the actual input and nominal input.}
\label{fig:01inputTrajectory}
\end{figure}

\begin{figure}
\centering
\includegraphics[width=.5\textwidth]{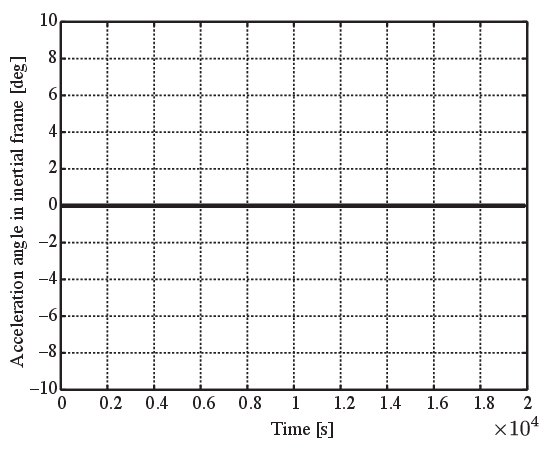}\protect\caption{Acceleration angle with respect to the $X$-axis in the inertial frame.}
\label{fig:01inputAngle}
\end{figure}

The first simulation aims at a formation reconfiguration under the
constraint on the input direction with respect to the inertial frame,
in which the terminated time is set to $t_{f}=2\pi/\Omega$. Figure
\ref{fig:01trajectory} shows the follower trajectory to the target
relative orbit, and the solid and dashed lines describe the follower
trajectory and the target orbit, respectively. The follower is relocated
to the target orbit and thus the proposed method successfully controls
the follower relative orbit. In Fig. \ref{fig:01inputTrajectory},
the input directions of the actual input and that of the nominal
input are shown. These input directions respectively describe the
circle and ellipse, and the input direction monotonically varies
at the same rate as the orbital rate of the leader. Figure \ref{fig:01inputAngle}
represents that the input angle with respect to the inertial frame
maintains at 0 deg throughout the maneuver, which means the input
direction keeps along with the desired direction. The cost function
is calculated as $1.78\times10^{-5}\,{\rm m^{2}/s^{3}}$, whereas
the cost function for the nominal optimal controller is $1.60\times10^{-5}\,{\rm m^{2}/s^{3}}$.
The energy consumption due to the additional input thus increases
11.2\%.

\begin{figure}
\centering
\includegraphics[width=.5\textwidth]{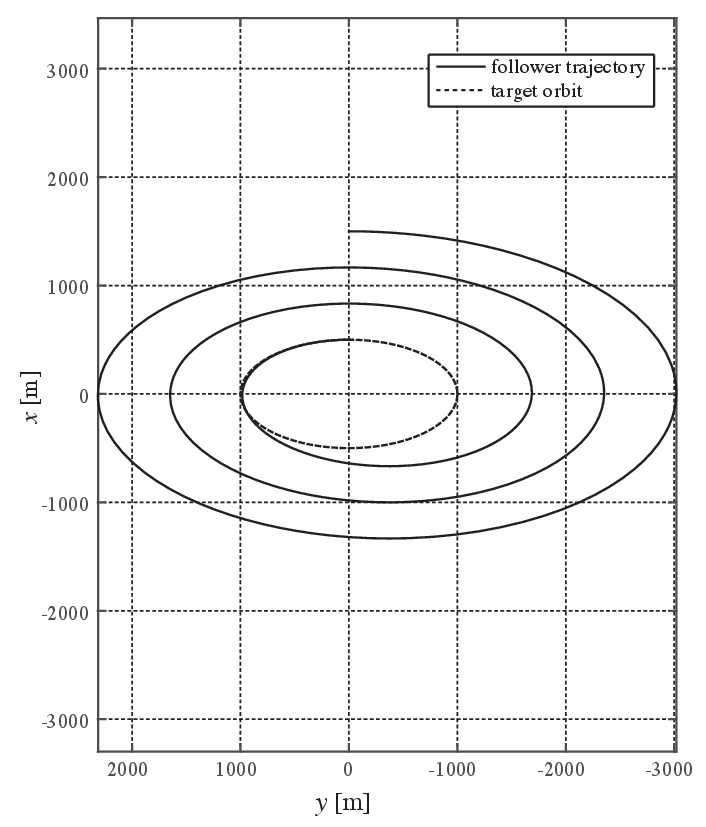}\protect\caption{Follower trajectory under the constraint on the input direction with
respect to the leader-fixed frame.}
\label{fig:02trajectory}
\end{figure}

\begin{figure}
\centering
\includegraphics[width=.5\textwidth]{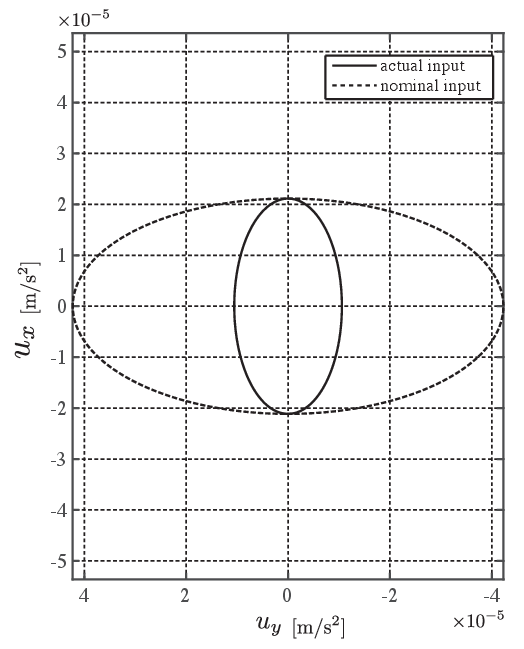}\protect\caption{Input directions of the actual input and nominal input.}
\label{fig:02inputTrajectory}
\end{figure}

The second simulation considers a formation control under the constraint
on the input direction with respect to the leader-fixed frame. The
initial and target relative semi-major axes are the same as the previous
simulation, and the terminated time and the forbidden region are set
to $t_{f}=6\pi/\Omega$ and $\bar{\delta}=30\,{\rm deg}$, respectively.
Figure \ref{fig:02trajectory} shows the follower trajectory to the
target orbit in which the solid line indicates the follower trajectory
and the dashed line means the target orbit. The relative orbit of
the follower is successfully controlled to the target one. In Fig.
\ref{fig:02inputTrajectory}, the solid line and dashed line represent
the actual input and nominal one, respectively. As discussed in the
previous section, the actual input direction becomes elliptic with
the ratio $\left|u_{x}\right|/\left|u_{y}\right|=2$ by superimposing
the additional input in Eq.~\eqref{eq:deltaX2}. 

\begin{figure}
\centering
\includegraphics[width=.5\textwidth]{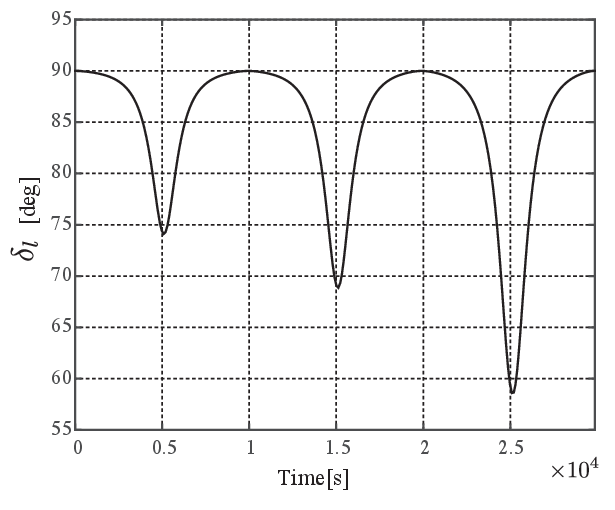}\protect\caption{Time history of the angle $\delta_{l}$.}
\label{fig:02inputAngle}
\end{figure}

Figure \ref{fig:02inputAngle} represents the time history of the angle $\delta_{l}$, and the angle
$\delta_{l}$ avoids being within the forbidden region throughout
the reconfiguration maneuver. 
If the maneuver time $t_{f}$ is too short, the follower satellite rapidly approaches the leader. 
This is not desirable for plume impingement avoidance because the thruster plume direction changes elliptically as shown in Fig.~\ref{fig:02inputTrajectory}. 
Therefore  when $\bar{\delta}$ is small, setting longer maneuver time $t_{f}$ will drive the follower to the leader gradually, resulting in the avoidance of the forbidden region. 
The limitation of the controller is $\bar{\delta}\approx90\,{\rm deg}$,
because the initial input angle is uniquely determined as $90\,{\rm deg}$
to realize the elliptic input direction. However, in practical situations,
at most $\bar{\delta}\leq60\,{\rm deg}$ would be reasonable, because
the plume impingement from a thruster tilted at $60\,{\rm deg}$ to
a surface is small enough (e.g., see \cite{Hyakutake:2000ip}). 

The cost function of this formation reconfiguration results in $8.34\times10^{-6}\,{\rm m^{2}/s^{3}}$,
whereas the cost function of the nominal controller is $5.34\times10^{-6}\,{\rm m^{2}/s^{3}}$.
Although the additional input increases $56\%$ of the energy consumption
and it is relatively large. One possible solution to solve this problem
is setting the longer terminated time $t_{f}$. In fact, the terminated
time in the second simulation is longer than the first one, which
results in the smaller cost function of the second simulation than
that of the first simulation.

\section{Conclusions}

This study has dealt with an optimal reconfiguration of satellites
under input directional constraints. The proposed method consists
two parts: 1) an energy optimal controller that guarantees a successful
formation reconfiguration and 2) additional inputs for satisfying
the input directional constraints. The optimal controller based on
the Fourier series is obtained without singularities, which enables
obtaining the condition for satisfying the constraints in analytical
forms. The additional inputs avoid the oscillation of the input direction
around a desired direction, and the input directional constraint with
respect to an inertial frame is satisfied. This method is modified
to accommodate the input directional constraint with respect to the
leader-fixed frame, in which a phase difference is compensated with
the additional inputs. Although these additional inputs affect the
transition of the follower, their error is compensated by virtual
target orbits. The advantage of the singularity avoidance also enables
analytically obtaining the virtual target orbits. Numerical simulations
verify the effectiveness of the proposed method for both constraints.
Possible extensions of this study are a formation control in an elliptic
orbit and under disturbances, which will be involved in future works.

\appendix{}
\section{Fourier Coefficients}

This appendix complements definitions and equations for the derived
optimal controller. The definitions of the variables in Eq.~\eqref{eq:integral}
are summarized as follows.

\begin{align}
&j_{0}  =-\int_{0}^{t_{f}}\frac{2}{\Omega}\mathrm{d}\tau\\
&j_{a}  =-\int_{0}^{t_{f}}\frac{2}{\Omega}\cos\left(\frac{2n\pi}{t_{f}}\right)\mathrm{d}\tau\\
&j_{b}  =-\int_{0}^{t_{f}}\frac{2}{\Omega}\sin\left(\frac{2n\pi}{t_{f}}\right)\mathrm{d}\tau
\end{align}
\begin{align}
&h_{0}  =-\int_{t_{0}}^{t_{f}}\cos\left(\Omega\tau\right)\mathrm{d}\tau\\
&h_{a}  =-\int_{0}^{t_{f}}\cos\left(\Omega\tau\right)\cos\left(\frac{2n\pi}{t_{f}}\right)\mathrm{d}\tau\\
&h_{b}  =-\int_{0}^{t_{f}}\cos\left(\Omega\tau\right)\sin\left(\frac{2n\pi}{t_{f}}\right)\mathrm{d}\tau
\end{align}
\begin{align}
l_{0} & =\int_{0}^{t_{f}}3\tau\mathrm{d}\tau\\
l_{a} & =\int_{0}^{t_{f}}3\tau\cos\left(\frac{2n\pi}{t_{f}}\right)\mathrm{d}\tau\\
l_{b} & =\int_{0}^{t_{f}}3\tau\sin\left(\frac{2n\pi}{t_{f}}\right)\mathrm{d}\tau
\end{align}
\begin{align}
k_{0} & =\int_{0}^{t_{f}}2\sin\left(\Omega\tau\right)\mathrm{d}\tau\\
k_{a} & =\int_{0}^{t_{f}}2\sin\left(\Omega\tau\right)\cos\left(\frac{2n\pi}{t_{f}}\right)\mathrm{d}\tau\\
k_{b} & =\int_{0}^{t_{f}}2\sin\left(\Omega\tau\right)\sin\left(\frac{2n\pi}{t_{f}}\right)\mathrm{d}\tau
\end{align}
\begin{align}
p_{0} & =\int_{0}^{t_{f}}\frac{1}{\Omega}\sin\left(\Omega\tau\right)\mathrm{d}\tau\\
p_{a} & =\int_{0}^{t_{f}}\frac{1}{\Omega}\sin\left(\Omega\tau\right)\cos\left(\frac{2n\pi}{t_{f}}\right)\mathrm{d}\tau\\
p_{b} & =\int_{0}^{t_{f}}\frac{1}{\Omega}\sin\left(\Omega\tau\right)\sin\left(\frac{2n\pi}{t_{f}}\right)\mathrm{d}\tau
\end{align}
\begin{align}
m_{0} & =\int_{0}^{t_{f}}2\Omega\cos\left(\Omega\tau\right)\mathrm{d}\tau\\
m_{a} & =\int_{0}^{t_{f}}2\Omega\cos\left(\Omega\tau\right)\cos\left(\frac{2n\pi}{t_{f}}\right)\mathrm{d}\tau\\
m_{b} & =\int_{0}^{t_{f}}2\Omega\cos\left(\Omega\tau\right)\sin\left(\frac{2n\pi}{t_{f}}\right)\mathrm{d}\tau
\end{align}
\begin{align}
q_{0} & =-\int_{0}^{t_{f}}\mathrm{d}\tau\\
q_{a} & =-\int_{0}^{t_{f}}\cos\left(\frac{2n\pi}{t_{f}}\right)\mathrm{d}\tau\\
q_{b} & =-\int_{0}^{t_{f}}\sin\left(\frac{2n\pi}{t_{f}}\right)\mathrm{d}\tau
\end{align}

The components of the matrix $A$ are also calculated as:
\begin{align*}
A_{11} & =j^{2}+l^{2}\\
A_{12} & =hj+kl\\
A_{13} & =lm+jp\\
A_{14} & =jq\\
A_{22} & =h^{2}+k^{2}\\
A_{23} & =km+hp\\
A_{24} & =kq\\
A_{33} & =m^{2}+p^{2}\\
A_{34} & =mq\\
A_{44} & =q^{2}
\end{align*}
where
\begin{align}
j^{2} & =\int_{0}^{t_{f}}\frac{4}{\Omega^{2}}\mathrm{d}\tau=\frac{4t_{f}^{2}}{\Omega^{2}}\\
l^{2} & =\int_{0}^{t_{f}}9\tau^{2}\mathrm{d}\tau=3t_{f}^{3}\\
h^{2} & =\int_{0}^{t_{f}}\cos^{2}\left(\Omega\tau\right)\mathrm{d}\tau=\frac{t_{f}}{2}+\frac{\sin\left(2\Omega t_{f}\right)}{4\Omega}\\
k^{2} & =\int_{0}^{t_{f}}4\sin^{2}\left(\Omega\tau\right)\mathrm{d}\tau=2t_{f}-\frac{\sin\left(2\Omega t_{f}\right)}{\Omega}\\
hj & =\int_{0}^{t_{f}}\frac{2}{\Omega}\cos\left(\Omega\tau\right)\mathrm{d}\tau=\frac{2\sin\left(\Omega t_{f}\right)}{\Omega^{2}}\\
kl & =\int_{0}^{t_{f}}6\tau\sin\left(\Omega\tau\right)\mathrm{d}\tau=\frac{6\left[\sin\left(\Omega t_{f}\right)-\Omega t_{f}\cos\left(\Omega t_{f}\right)\right]}{\Omega^{2}}\\
lm & =\int_{0}^{t_{f}}\frac{6\tau\cos\left(\Omega\tau\right)}{\Omega}\mathrm{d}\tau=\frac{6\left[-1+\cos\left(\Omega t_{f}\right)+\Omega t_{f}\sin\left(\Omega t_{f}\right)\right]}{\Omega^{3}}\\
jp & =-\int_{0}^{t_{f}}\frac{2}{\Omega^{2}}\sin\left(\Omega\tau\right)\mathrm{d}\tau=\frac{2\left[-1+\cos\left(\Omega t_{f}\right)\right]}{\Omega^{3}}\\
lq & =-\int_{0}^{t_{f}}3\tau\mathrm{d}\tau=-\frac{3t_{f}^{2}}{2}\\
km & =\int_{0}^{t_{f}}\frac{4\sin\left(\Omega\tau\right)\cos\left(\Omega\tau\right)}{\Omega}\mathrm{d}\tau=\frac{2\sin^{2}\left(\Omega t_{f}\right)}{\Omega^{2}}\\
hp & =-\int_{0}^{t_{f}}\sin\left(\Omega\tau\right)\cos\left(\Omega\tau\right)/\Omega\mathrm{d}\tau=-\frac{\sin^{2}\left(\Omega t_{f}\right)}{2\Omega^{2}}\\
kq & =-\int_{0}^{t_{f}}2\sin\left(\Omega\tau\right)\mathrm{d}\tau=\frac{2\left[-1+\cos\left(\Omega t_{f}\right)\right]}{\Omega}\\
mq & =-\int_{0}^{t_{f}}2\cos\left(\Omega\tau\right)/\Omega\mathrm{d}\tau=-\frac{2\sin^{2}\left(\Omega t_{f}\right)}{\Omega^{2}}
\end{align}
The analytical form of the matrix $A$ can be simplified setting $t_{f}=2N\pi/\Omega$,
and the Lagrange multipliers are calculated from Eqs.~\eqref{eq:givenQ}
and~\eqref{eq:Q} as follows.
\begin{equation}
\bm{\lambda}=\left[\begin{array}{cccc}
\frac{15n^{3}}{15N^{2}\pi^{2}-52} & \frac{11n^{2}}{52N\pi-15N^{3}\pi^{3}} & \frac{5n^{3}}{30N^{3}\pi^{3}-104N\pi} & \frac{15n^{2}}{30N^{2}\pi^{2}-104}\\
\frac{36n^{2}}{15N^{2}\pi^{2}-52} & \frac{n\left(3\pi^{2}N^{2}+16\right)}{52N\pi-15N^{3}\pi^{3}} & \frac{6n^{2}}{15N^{3}\pi^{3}-52N\pi} & \frac{18n}{15N^{2}\pi^{2}-52}\\
\frac{3n^{3}}{5N\pi} & 0 & 0 & \frac{2n^{2}}{5N\pi}\\
\frac{30N^{2}\pi^{2}n^{2}+52n^{2}}{15N^{3}\pi^{3}-52N\pi} & \frac{33n}{52-15N^{2}\pi^{2}} & \frac{15n^{2}}{30N^{2}\pi^{2}-104} & \frac{15n\pi^{2}N^{2}+26n}{15N^{3}\pi^{3}-52N\pi}
\end{array}\right]\left[\bm{x}\left(t_{f}\right)-\bm{x}_{h}\right]\label{eq:LagrangeMultipliers}
\end{equation}

\bibliographystyle{elsarticle-harv}
\biboptions{authoryear}
\bibliography{yoshimura}


%


\end{document}